\documentclass[prb,11pt,onecolumn,floatfix]{revtex4}
\usepackage{graphicx,amssymb,amsmath,pdfpages}

\begin{document}

\title{Ultrafast Response of Monolayer Molybdenum Disulfide Photodetectors}

\author{Haining Wang,Changjian Zhang,Weimin Chan,Sandip Tiwari,Farhan Rana}
\affiliation{School of Electrical and Computer Engineering, Cornell University, Ithaca, NY, USA}
\email{hw343@cornell.edu}

\begin{abstract}

The strong light emission and absorption exhibited by single atomic layer transitional metal dichalcogenides in the visible to near-infrared wavelength range makes them attractive for optoelectronic applications. In this work, using two-pulse photovoltage correlation technique, we show that monolayer molybdenum disulfide photodetector can have intrinsic response times as short as 3 ps implying photodetection bandwidths as wide as 300 GHz. The fast photodetector response is a result of the short electron-hole and exciton lifetimes in this material. Recombination of photoexcited carriers in most two-dimensional metal dichalcogenides is dominated by non-radiative processes, most notable among which is Auger scattering. The fast response time, and the ease of fabrication of these devices, make them interesting for low-cost ultrafast optical communication links.      
 
\end{abstract}

\maketitle

\begin{figure}[tbp]
  \centering
  \includegraphics[width=0.8\textwidth]{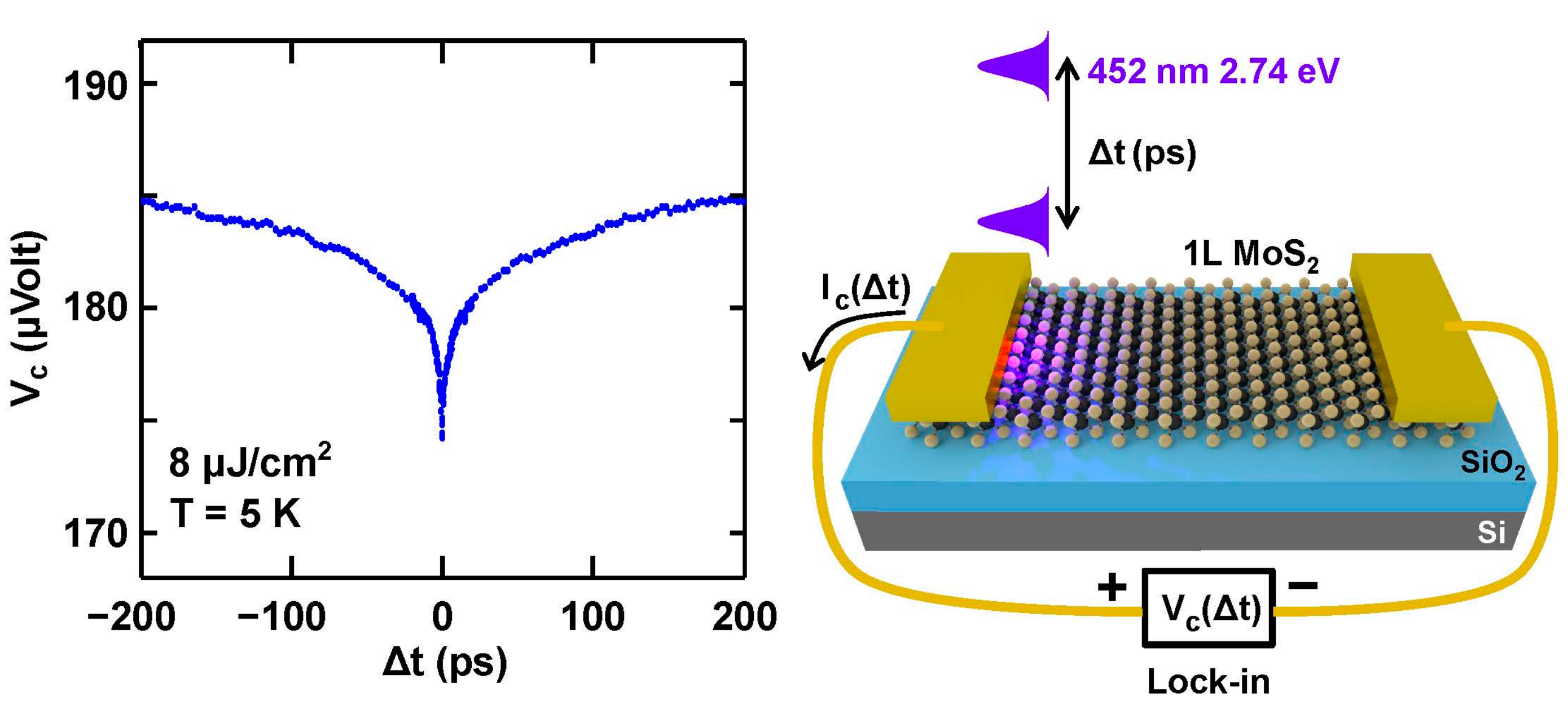}
  \caption
      {\textbf{Two-Pulse Photovoltage Correlation Experiment.}} 
  \label{fig:cover}
\end{figure}

Two dimensional (2D) transition metal dichalcogenides (TMDs) have emerged as interesting materials for low-cost opto-electronic devices, including photodetectors, light-emitting diodes, and, more recently, lasers~\cite{Mak10,Splendiani10,Wang12,Mak13,Lopez13,Ross14,Hua12,Steiner13,Baugher14,Changjian14,Sanfeng15}. In the case of photodetectors, the response time and the quantum efficiency are two important figures of merit. The intrinsic response time of TMD photodetectors and the ultimate limits on the speed of operation are unknown. The reported quantum efficiencies of TMD materials and devices are typically in the .0001-0.01 range~\cite{Lopez13,Hua12,Wang15,Steiner13,Ross14,Grossman14,Thomas14,Baugher14,Thomas14b}, indicating that most of the electrically injected or optically generated electrons and holes recombine nonradiatively. Understanding the nonradiative carrier recombination mechanisms, as well as the associated time scales, is therefore important. Previously, ultrafast optical/THz pump-probe as well as ultrafast photoluminescence techniques have been used, by the authors and others, to study the ultrafast carrier dynamics in metal dichalcogenides and in molybdenum disulfide (MoS$_{2}$) in particular~\cite{Wang15,Huang13,WangR12,Choi13,Sun14,Lagarde14,Schuller11,Docherty14}. In these measurements, free-carrier recombination dynamics, exciton formation and recombination dynamics, refractive index changes, optical/THz intraband and interband conductivity changes, as well as the dynamics associated with carriers trapped in optically active midgap defects are all expected to play a role to varying degrees and, consequently, the results have been difficult to interpret and reconcile.

In this letter, we present experimental results on ultrafast two-pulse photovoltage correlation (TPPC) measurements on monolayer MoS$_2$ metal-semiconductor photodetectors. In TPPC measurements, a photodetector is excited with two identical optical pulses separated by a time delay and the integrated detector photoresponse (either photovoltage or photocurrent response) is recorded as a function of the time delay. TPPC thus uses the photodetector to perform an optical correlation measurement. The nonlinearity of the photoresponse with respect to the optical pulse energy enables one to determine ultrafast intrinsic temporal response of the detector with sub-picosecond resolution~\cite{Ulrich11,Holleitner11,Holleitner12,McEuen13,Sun12}. Our measurements show that the photovoltage is suppressed when the two optical pulses arrive together indicating a saturation of the photoresponse. As the time delay between the two pulses is increased from zero, the photovoltage recovers, and the recovery, as a function of the time delay, exhibits two distinct timescales: (i) a fast timescale of the order of 3 to 5 ps, and (ii) a slow timescale of around 80 to 110 ps. These two timescales are found to be largely independent of the temperature, exhibits only a mild dependence on the pump fluence, and varies a little from sample to sample. Between 50$\%$-75$\%$ of the photovoltage correlation response recovers on the fast timescale implying that ultrafast TMD photodetectors with (8 dB) current modulation bandwidths in the 200-300 GHz range are possible. The fast response speed is a result of the short lifetime of the photoexcited carriers. Since TPPC measures the photovoltage (or photocurrent), this technique is sensitive only to the total photoexcited carrier population, including both bound (excitons) and free carriers, that contributes to the photoresponse. TPPC therefore also offers important and unique insights into the carrier recombination dynamics. The temperature and pump fluence dependence of our TPPC results are consistent with defect-assisted recombination as being the dominant mechanism, in which the the photoexcited electrons and holes, both free and bound (excitons), are captured by defects via Auger scattering~\cite{Wang15b}. Strong Coulomb interactions in 2D TMDs, including the correlations in the positions of free and bound electrons and holes arising from the attractive interactions, result in large carrier capture rates by defects via Auger scattering~\cite{Wang15b}. Our results underscore the trade-off between speed and quantum efficiency in TMD photodetectors.

\section{Results}
\subsection{Two-Pulse Photovoltage Correlation Technique} 

\begin{figure}[tbp]
  \centering
  \includegraphics[width=0.8\textwidth]{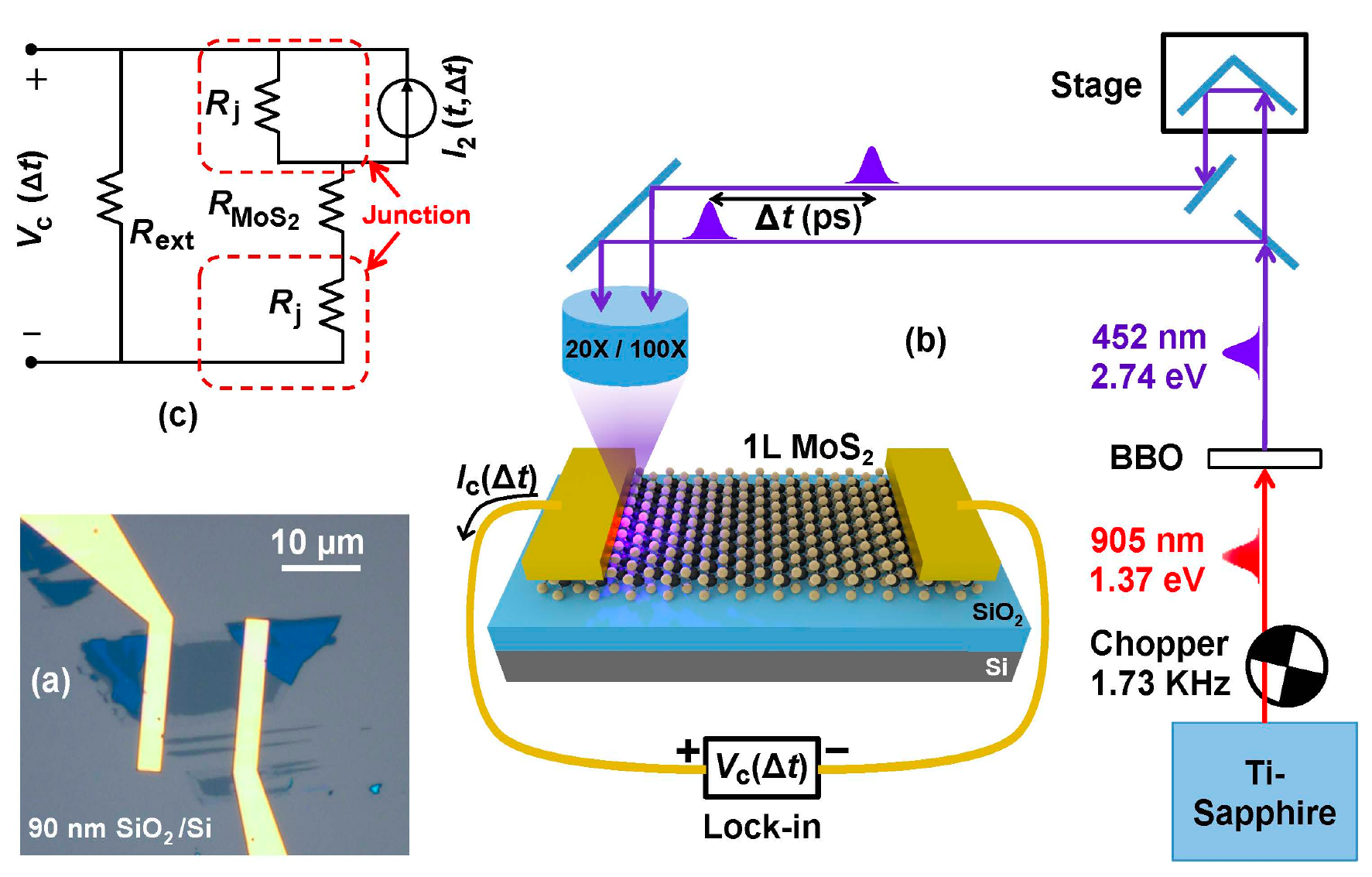}
  \caption
      {\textbf{TPPC experiment and circuit model of metal-MoS$_2$ photodetector.} (a) Optical micrograph of a fabricated back-gated monolayer metal-MoS$_2$ photodetector on SiO$_2$/Si substrate is shown. (b) A schematic of the two-pulse photovoltage correlation (TPPC) experiment is shown. Two time-delayed 452 nm optical pulses, both obtained via upconversion from a single Ti:Sapphire laser, are focused at one of the metal-semiconductor Schottky junctions. The generated DC photovoltage is recorded as a function of the time delay between the pulses. A lock-in detection scheme is used to improve the signal-to-noise ratio. The arrow indicates the positive direction of the photocurrent (and the sign of the measured photovoltage) form the illuminated metal contact. (c) A low-frequency circuit model of the device and measurement. The current source $I_{2}(t,\Delta t)$ represents the short circuit current response of the junction in response to two optical pulses separated by time $\Delta t$. $R\mathrm{_{j}}$ is the resistance of the metal-MoS$_{2}$ junction. $R\mathrm{_{MoS_{2}}}$ is the resistance of the MoS$_{2}$ layer. $R\mathrm{_{ext}}$ is the external circuit resistance (including the $\sim$10 M$\Omega$ input resistance of the measurement instrument).} 
  \label{fig:device}
\end{figure}

Microscope image of a monolayer metal-MoS$_2$ photodetector is shown in Figure \ref{fig:device}(a), and the schematic in Figure \ref{fig:device}(b) depicts the setup for a two-pulse photocurrent/photovoltage correlation (TPPC) experiment. A $\sim$80 fs, 905 nm (1.37 eV) center wavelength, optical pulse from a $\sim$83 MHz repetition rate Ti-Sapphire laser is frequency doubled to 452 nm (2.74 eV, $\sim$150 fs) by a beta-BaB$_{2}$O$_{4}$ crystal, then mechanically chopped at 1.73 KHz, and then split into two pulses by a 50/50 beam splitter. The time delay $\Delta t$ between these two pulses is controlled by a linear translation stage. The resulting voltage across the photodetector is measured as a function of the time delay between the pulses using a lock-in amplifier with a 10 M$\Omega$ input resistance. In experiments, the maximum photoresponse was obtained when the light was focused on the sample near one of the metal contacts of the device, and the photoresponse decayed rapidly as the center of the focus spot was moved more than half a micron away from the metal contact.  The direction of the DC photocurrent, and the resulting sign of the measured DC photovoltage are shown in Figure \ref{fig:device}(b), and were determined without using the lock-in. Photovoltage was always positive at the contact near which the light was focused. Figure \ref{fig:device}(c) shows a low-frequency circuit model of the device and the measurement. The circuit model shown can be derived from a high-frequency circuit model(Supplementary Figure 2 and Note 3). If the time-dependent short circuit current response of the illuminated junction to a single optical pulse is  $I_{1}(t)$, and to two optical pulses separated by time $\Delta t$ is $I_{2}(t,\Delta t)$, and the external resistance $R\mathrm{_{ext}}$ is much larger than the total device resistance, then the measured DC voltage $V\mathrm{_{c}}(\Delta t)$ is approximately equal to $(R\mathrm{_{j}}/T\mathrm{_{R}}) \int I_{2}(t,\Delta t) \, dt$, where $T\mathrm{_{R}}$ is the pulse repetition period, $R\mathrm{_{j}}$ is the resistance of the metal-MoS$_{2}$ junction, and the time integral is over one complete period. As the time delay $\Delta t$ becomes much longer than the duration of $I_{1}(t)$, one expects $V\mathrm{_{c}}(\Delta t)$ to approach $(2R\mathrm{_{j}}/T\mathrm{_{R}}) \int I_{1}(t) \, dt$.

\subsection{Experimental Results}

\begin{figure}[tbp]
  \centering
   \includegraphics[width=1.0\textwidth]{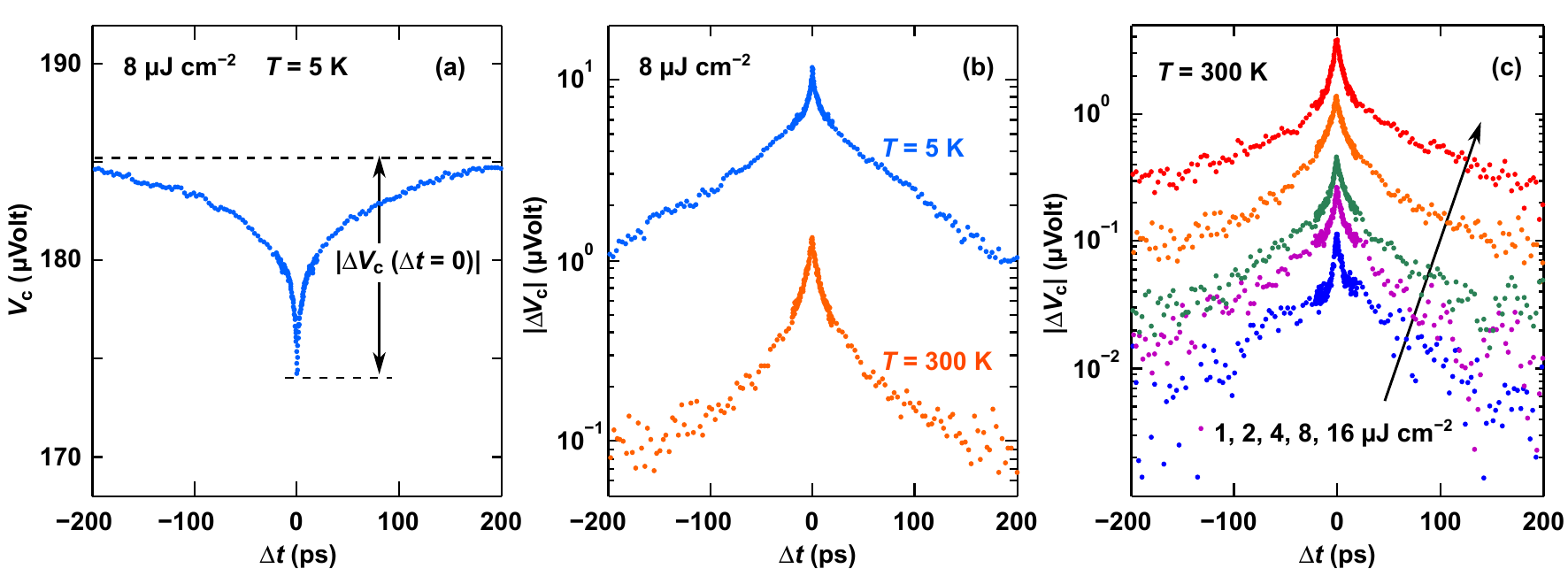}
  \caption
      {\textbf{TPPC experiment results.} (a) The measured two-pulse photovoltage correlation (TPPC) signal $V\mathrm{_{c}}(\Delta t)$ is plotted as a function of the time delay $\Delta t$ between the pulses. $T$ = 5K. Gate bias is 0 V. The pump fluence is 8 $\mu$J cm$^{-2}$. The quantity $\Delta V\mathrm{_{c}}(\Delta t)$ is the difference between $V\mathrm{_{c}}(\Delta t)$ and its maximum value which occurs when $\Delta t \rightarrow \infty$. (b) $|\Delta V\mathrm{_{c}}(\Delta t)|$ is plotted on a log scale as a function of the time delay $\Delta t$ between the pulses in order to show the two distinct timescales exhibited by  $V\mathrm{_{c}}(\Delta t)$. The two curves are for two different extreme temperatures: $T$= 5 K and $T$ = 300 K. The plot shows two distinct timescales: (i) a fast timescale of $\sim$4.3 ps, and (ii) a slow timescale of $\sim$105 ps. Gate bias is 0 V. The pump fluence is 8 $\mu$J cm$^{-2}$. The two different timescales are observed at both temperatures and these timescales are largely temperature independent. (c) The measured two-pulse photovoltage correlation (TPPC) signal $|\Delta V\mathrm{_{c}}(\Delta t)|$ is plotted as a function of the time delay $\Delta t$ between the pulses for different pulse fluences:  1, 2, 4, 8, and 16 $\mu$J cm$^{-2}$. $T$ = 300K. Gate bias is 0 V. The two different timescales are observed at all values of the pump fluence and these timescales are not very sensitive to the pump fluence.}
  \label{fig:Data}
\end{figure}

\ref{fig:Data}(a) shows the measured two-pulse photovoltage correlation signal $V\mathrm{_{c}}(\Delta t)$ plotted as a function of the time delay $\Delta t$ between the pulses. The substrate temperature is 5K, the gate bias is 0 V, and the pump fluence is 8 $\mu$J cm$^{-2}$.  $V\mathrm{_{c}}(\Delta t)$ is minimum when the two pulses completely overlap in time (i.e. when $\Delta t =0$). This implies, not surprisingly, that the photovoltage response of the detector to an optical pulse is a sublinear function of the optical pulse energy. As $\Delta t$ increases from zero, $V\mathrm{_{c}}(\Delta t)$ also increases from its minimum value at $\Delta t =0$. As $\Delta t$ becomes much longer than the duration of the response transient of the detector to an optical pulse, $V\mathrm{_{c}}(\Delta t)$ approaches a constant value. The timescales over which $V\mathrm{_{c}}(\Delta t)$ goes to the constant value are related to the timescales associated with the response transient of the detector to an optical pulse. These timescales are better observed in the measured data if the magnitude of $\Delta V\mathrm{_{c}}(\Delta t)$, defined as $V\mathrm{_{c}}(\Delta t) - V\mathrm{_{c}}(\infty)$, is plotted on a log scale, as shown in Figure \ref{fig:Data}(b). The plot in Figure \ref{fig:Data}(b) shows two distinct timescales: (i) a fast timescale of $\sim$4.3 ps, and (ii) a slow timescale of $\sim$105 ps. In different devices, the fast timescales were found to be in the 3-5 ps range, and the slow timescales were in the 90-110 ps range. The fast timescales imply (8 dB) current modulation bandwidths wider than 300 GHz.  

Measurements were performed at different temperatures and using different pump pulse fluences in order to understand the mechanisms behind the photoresponse and the associated dynamics. Figure \ref{fig:Data}(b) shows $|\Delta V\mathrm{_{c}}(\Delta t)|$ plotted for two different extreme temperatures: $T$ = 5K and $T$ = 300K. Gate bias is 0 V. The pump fluence is 8 $\mu$J cm$^{-2}$. Two distinct timescales are observed at both temperatures and these timescales are largely independent of the temperature. Measurements performed at intermediate temperatures provided the same results. $|\Delta V\mathrm{_{c}}(\Delta t)|$ was found to be larger at smaller temperatures. We attribute this to the increase in the metal-semiconductor junction resistance $R\mathrm{_{j}}$ at lower temperatures. Figure \ref{fig:Data}(c) shows $|\Delta V\mathrm{_{c}}(\Delta t)|$ plotted for different values of the pump pulse fluence. The pump pulse fluence was varied from 1 $\mu$J cm$^{-2}$ to 16 $\mu$J cm$^{-2}$. Higher values of the pump fluence were not used in order to avoid optical damage to the sample~\cite{Wang15}. $T$ = 300K and gate bias is 0 V. Figure \ref{fig:Data}(c) shows that the same two timescales are observed for all values of the pump fluence and these timescales are not very sensitive to the pump fluence. The observed timescales also did not change in any significant way under a positive or a negative gate bias (Supplementary Figure 3, Figure 4 and Note 4). The internal and external detector quantum efficiencies were estimated from the measured values of $V\mathrm{_{c}}(\infty)$ and the junction resistance $R\mathrm{_{j}}$ to be in the 0.008-0.016 and 0.001-0.002 ranges, respectively. 

The two different timescales observed in our two-pulse photovoltage correlation experiments match well with the two different timescales observed previously in ultrafast optical/THz pump-probe studies of carrier dynamics as well as in ultrafast photoluminescence studies of MoS$_{2}$ monolayers~\cite{Wang15,Docherty14}. It is therefore intriguing if the model for defect-assisted carrier recombination reported previously by the authors~\cite{Wang15,Wang15b} for explaining the ultrafast carrier dynamics in MoS$_{2}$ monolayers can be used to obtain photovoltage correlations that are in good agreement with the experimental results reported in this letter. We show below that this is indeed the case.

\subsection{Ultrafast Photoresponse of the Metal-MoS$_2$ Junction}
 
Understanding the ultrafast photoresponse of the detector, and in particular the short circuit current response $I_{2}(t,\Delta t)$, is important for interpreting the experimental results. Figure \ref{fig:fitting}(a) depicts the band diagram of the metal-MoS$_{2}$ junction (plotted in the plane of the MoS$_{2}$ layer) after photoexcitation with an optical pulse. Given the Schottky barrier height of 100-300 meV~\cite{Steiner13}, the width of the MoS$_{2}$ region near the metal with a non-zero lateral electric field is estimated to be to $\sim$100-300 nm~\cite{Shik89}. As a result of light diffraction from the edge of the metal contact, light scattering from the substrate, and plasmonic guidance, a portion of the MoS$_{2}$ layer of length equal to a few hundred nanometers is photoexcited even underneath the metal (Supplementary Figure 1 and Note 1).  The photoresponse of graphene photodetectors has been explained in terms of contributions from photovoltaic and photothermoelectric contributions~\cite{Ulrich11,Holleitner11,Holleitner12,McEuen13,Sun12,Koppens15}. In our MoS$_{2}$ samples, the carrier mobilities and diffusivities are 2-3 orders of magnitude smaller than in graphene and the time period in which most of the photoexcited carriers recombine and/or are captured by defects is in the few picoseconds range~\cite{Wang15,Wang15b,Docherty14,Lagarde14,Schuller11}. Assuming similar mobilities and diffusivities of electrons and holes in MoS$_{2}$, the photoexcited carriers, both free and bound (excitons), move, either by drift in the junction lateral electric field or by diffusion, less than $\sim$10 nm in 5 ps before they recombine and/or are captured by defects. The photoexcited carrier distributions therefore do not change significantly in space during their lifetime. Separation of electrons and holes either by the junction lateral electric field or at the metal-MoS$_{2}$ interface will contribute to the integral $\int I_{2}(t,\Delta t) \, dt \, (\propto V\mathrm{_{c}}(\Delta t))$ (the measured dependence of the photoresponse on the junction electric field and the gate voltage is discussed in the Supplementary Figure 3, Figure 4 and Note 4. We assume that the integral $\int I_{2}(t,\Delta t) \, dt$ is approximately proportional to the integral $\int \left[ p'(t,\Delta t) + n'(t,\Delta t) \right] \, dt$ (Supplementary Note 2). Here, $p'(t,\Delta t)$ and $n'(t,\Delta t)$ are the time-dependent photoexcited electron and hole densities in the junction, including carriers both free and bound (excitons). This assumption, although simple, allows one to relate the measured photoresponse to the carrier dynamics and, as shown below, the results thus obtained are in excellent agreement with the experiments. We expect that on much longer timescales, when the photoexcited carriers have recombined or been captured, the photoresponse is entirely thermoelectric in nature, as is the case in metal-graphene photodetectors~\cite{Ulrich11,Holleitner12,McEuen13,Sun12,Koppens15}. But in this letter we focus on the dynamics occurring on only short timescales.

\subsection{Carrier Capture/Recombination Model and Comparison with Data}
 
It is known that most of the photoexcited carriers in monolayer MoS$_{2}$ recombine non-radiatively~\cite{Wang15,Wang15b,Thomas14b}. The temperature independence as well as the sample dependence of the recombination rates in previously reported works suggested that free and bound (excitons) carriers recombine via capture by defects through Auger scattering~\cite{Wang15,Wang15b}. Monolayer MoS$_2$ can have several different kinds of defects, including grain boundaries, line defects, interstitials, dislocations and vacancies~\cite{Sofo04,Komsa12,Seifert13,Zhou13,Kim14,Guinea14,Robertson13}. The strong electron and hole Coulomb interaction in monolayer TMDs makes defect capture via Auger scattering very effective~\cite{Wang15b}. The time constants observed in this work in the photovoltage correlations are also temperature independent (see Figure \ref{fig:Data}(b)) and also match well with the time constants observed previously in optical/THz pump-probe and photoluminescence measurements~\cite{Wang15,Docherty14}. We therefore use the model for carrier capture by defects via Auger scattering in MoS$_{2}$ presented by Wang et~al.~\cite{Wang15,Wang15b} to model our TPPC experimental results (Supplementary Figure 5 and Note 5). The model assumes carrier capture by two different defect levels, one fast and one slow~\cite{Wang15}. The essential dynamics captured by the model, and their relationships to the experimental observations, are as follows~\cite{Wang15}. After photoexcitation, electrons and holes thermalize and lose most of their energy on a timescale much shorter than our experimental resolution ($\sim$0.5-1.0 ps) and, therefore, thermalization is assumed to happen instantly in our model~\cite{Kaas14}. Most of the photoexcited holes (both free and bound), followed by most of the electrons, are captured by the fast defects within the first few picoseconds after photoexcitation. During the same period, a small fraction of the holes is also captured by the slow defects. This rapid capture of the photoexcited electrons and holes is responsible for the fast time constant observed in our photovoltage correlation experiments. The remaining photoexcited electrons are captured by the slow defects on a timescale of 65-80 ps and this slow capture of electrons is responsible for the slow time constant observed in our photovoltage correlation experiments. We should point out here that the two time constants observed in the photovoltage correlation signal $\Delta V\mathrm{_{c}}(\Delta t)$ are always slightly longer than the corresponding two time constants exhibited by the carrier densities in direct optical pump-probe and measurements~\cite{Wang15}, as is to be expected in the case of correlation measurements. Finally, the superlinear dependence of the carrier capture rates on the photoexcited carrier densities in the model, and therefore on the optical pulse energy, results in the experimentally observed negative value of the photovoltage correlation signal $\Delta V\mathrm{_{c}}(\Delta t = 0)$ at zero time delay. The values of the fitting parameters used in the theoretical model to fit the experimental data are given in the Supplementary Table 1 and are almost identical to the values extracted previously from direct optical pump-probe measurements of the carrier dynamics in monolayer MoS$_{2}$~\cite{Wang15}.

\begin{figure}[tbp]
  \centering
   \includegraphics[width=1.0\textwidth]{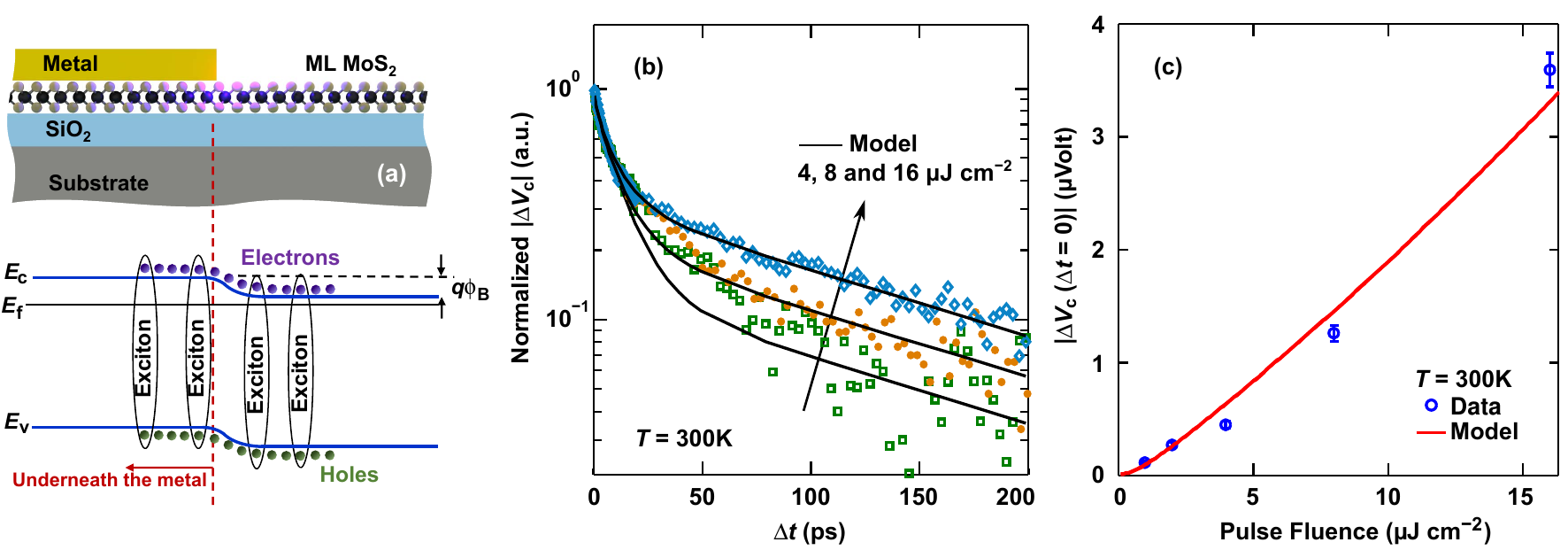}
  \caption
      {\textbf{Theoretical modelling and fitting of TPPC experiment results.} (a) The energy band diagram (bottom) of the the metal-MoS$_{2}$ junction (top) is plotted as a function of the position in the plane of the MoS$_{2}$ monolayer after photoexcitation with an optical pulse. The Schottky barrier height is $\phi\mathrm{_{B}}$. The Figure is not drawn to scale. (b) The measured (symbols) and computed (solid lines) photovoltage correlation signals $|\Delta V\mathrm{_{c}}(\Delta t)|$, normalized to the maximum value, are plotted as a function of the time delay $\Delta t$ for different pump fluence values: 4, 8, and 16 $\mu$J cm$^{-2}$. $T$ = 300K. The carrier capture model reproduces all the timescales observed in the measurements over the entire range of the pump fluence values used. (c) The scaling of the measured (symbols) and computed (solid line) values of $|\Delta V\mathrm{_{c}}(\Delta t = 0)|$ with the pump pulse fluence is shown. The error bars are the recorded peak-to-peak noise level of lock-in amplifier during each measurement.}
  \label{fig:fitting}
\end{figure}

The comparison between the data and the model are shown in Figure \ref{fig:fitting}(b) which plots the measured and computed photovoltage correlation $|\Delta V\mathrm{_{c}}(\Delta t)|$ as a function of the time delay $\Delta t$ for different pump fluence values: 4, 8, and 16 $\mu$J cm$^{-2}$. All curves are normalized to a maximum value of unity (since the model gives the photovoltage correlation signal up to a multiplicative constant). The model not only reproduces the very different timescales observed in $|\Delta V\mathrm{_{c}}(\Delta t)|$ measurements, it achieves a very good agreement with the data over the entire range of the pump fluence values used in our experiments for the same values of the parameters (Supplementary Table 1). Figure \ref{fig:fitting}(c) shows the scaling of the measured and computed values of $|\Delta V\mathrm{_{c}}(\Delta t = 0)|$ with the pump pulse fluence. Again, a very good agreement is observed between the model and the data.

\section{Discussion}

Our results reveal the fast response time and the wide bandwidth of metal-MoS$_{2}$ photodetectors and show that these detectors can be used for ultrafast applications. Our results also shed light on the carrier recombination mechanisms and the associated timescales. Although we focused mainly on the carrier dynamics in this paper, the device intrinsic resistances and capacitances are not expected to fundamentally limit the device speed because of the rather small capacitances associated with the lateral metal-semiconductor junctions (Supplementary Note 3). An obstacle to using TMDs in practical light emission and detection applications is the small values of the reported quantum efficiencies in these materials~\cite{Lopez13,Hua12,Wang15,Steiner13,Ross14,Grossman14,Thomas14,Baugher14,Thomas14b}. In photodetectors, the response speed and the quantum efficiency are often inversely related~\cite{Donati99}. In most semiconductor photovoltaic detectors with large carrier mobilities and long recombination times (e.g. group III-V semiconductor photodetectors~\cite{Donati99}), the transit time of the photogenerated carriers through the junction depletion region determines the detector bandwidth~\cite{Donati99}. Given the relatively small carrier mobilities and diffusivities in MoS$_{2}$, the fast carrier recombination times determine the speed in our metal-MoS$_{2}$ detectors. The price paid for the the fast response time is the small internal quantum efficiency: most of the photogenerated carriers recombine before they make it out into the circuit. The best reported carrier mobilities in monolayer MoS$_{2}$ are an order of magnitude larger than in our devices and, therefore, metal-MoS$_{2}$ photodetectors with internal quantum efficiencies around 0.1 (approximately an order of magnitude larger than of our devices) are possible without sacrificing the wide bandwidth. Density of defects, which contribute to carrier trapping and recombination, is also a parameter that can be potentially controlled in 2D TMD optoelectronic devices to meet the requirements for ultrafast or high quantum efficiency applications. In addition, vertical heterostructures of 2D TMD materials can also be used to circumvent the transport bottleneck in high speed applications~\cite{Koppens15b}. 

\section{Methods}
\subsection{Device Fabrication and Characterization}
Monolayer MoS$_2$ samples were mechanically exfoliated from bulk MoS$_2$ crystal (obtained from 2D Semiconductors Inc.) and transferred onto highly n-doped Si substrates with 90 nm thermal oxide. Monolayer sample thickness was confirmed by Raman and reflection spectroscopies~\cite{Ryu10}. Au metal contacts (with a very thin Cr adhesion layer) were patterned and deposited onto the samples using electron-beam lithography. The doped Si substrate acted as the back gate. Microscope image of a 10$\times$10 $\mu$m$^{2}$ area device is shown in Figure \ref{fig:device}(a). Carrier densities and mobilities in the devices were determined using electrical transport measurements on devices of different dimensions. The devices were found to be n-doped with electron densities around $1\times 10^{12}$-$2\times 10^{12}$ cm$^{-2}$ (under zero gate bias). The intrinsic doping was attributed to impurities and defect levels~\cite{Steiner13}. The electron mobility in the devices was found to be in the $15-20$ cm$^{2}$ V$^{-1}$s$^{-1}$ range at 5K. The zero gate bias device resistance was typically less than 1 M$\Omega$ at all temperatures for a 10$\times$10 $\mu$m$^{2}$ area device. While the device resistance decreased under a positive gate bias, no signature of hole conduction was observed even when a large negative gate bias was applied indicating that the Fermi level in the MoS$_{2}$ layer was likely pinned at defect levels within the bandgap under a negative gate bias. The devices were mounted in a helium-flow cryostat and the temperature was varied between 5K and 300K during measurements. The zero gate bias device resistance was found to be a function of the temperature and decreased almost linearly with the temperature from $\sim$1 M$\Omega$ at 5K to values 5-7 times smaller at 300K. The total device resistance was dominated by the metal Schottky contacts to the device. For example, at 5K the resistance contributed by the 10$\times$10 $\mu$m$^{2}$ area MoS$_{2}$ strip is estimated to be in the 0.10-0.20 M$\Omega$ range (from the measured mobility values), which is approximately only one-tenth to one-fifth of the total device resistance. The reported Schottky barrier heights between similarly n-doped monolayer MoS$_2$ and Au/Cr contacts are in the 100-300 meV range~\cite{Steiner13}. Depending on how strongly the Fermi level gets pinned at the defect levels in the bandgap, the lateral potential drop in the MoS$_{2}$ layer at the metal-MoS$_{2}$ interface (depicted in Figure \ref{fig:fitting}(a)) could be equal to or smaller than the Schottky barrier height. 
\subsection{TPPC Experiment Setup}
In the TPPC measurements, the two 452 nm optical pulses were cross-polarized to minimize interference and focused onto the device using a 20X or a 100X microscope objective resulting in a minimum focus spot size of around 0.75 $\mu$m. Optical absorption in monolayer MoS$_{2}$ layers was characterized using a confocal microscope based relection/transmission setup~\cite{Wang15} and yielded around 11\%-12\% absorption (single-pass) in monolayer MoS$_{2}$ on oxide at 452 nm (pump pulse wavelength). Measurement of the photovoltage using a high input impedance voltage amplifier (Lock-in in our case) was found to give a much better signal-to-noise ratio than the measurement of the photocurrent directly using a low input impedance transimpedance amplifier. 

\hfill \\
\noindent
\textbf{Acknowledgments}
\newline
The authors would like to acknowledge helpful discussions with Jared H. Strait, Michael G. Spencer, and Paul L. McEuen, and support from CCMR under NSF grant number DMR-1120296, AFOSR-MURI under grant number FA9550-09-1-0705, ONR under grant number N00014-12-1-0072, and the Cornell Center for Nanoscale Systems funded by NSF.

\hfill \\
\noindent
\textbf{Author contributions}
\newline
H. W. performed the experiments, modelling, and simulation. C.Z and W.M.C fabricated the devices. S.T. and F.R initiated and supervised the work.

\hfill \\
\noindent
\textbf{Additional information}
\newline
Correspondence and requests for materials should be addressed to H.W and F.R.

\newpage
\clearpage


\section*{Supplementary material}

**Supplementary Note 1: Numerical Simulation of the Photoexcitation of the Metal-MoS$_2$ Junction**

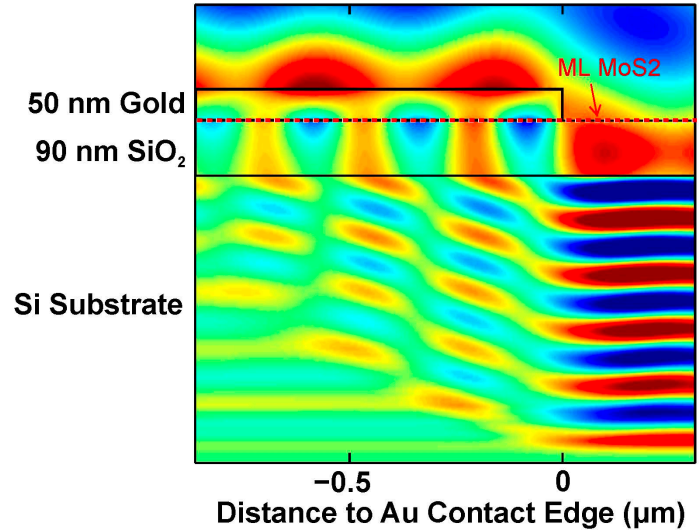

**Supplementary Figure** 1: **Numerical simulation of photoexcitation at metal-MoS$_2$ junction.** Finite Difference Time domain (FDTD) simulation of the optical excitation of the metal-MoS$_2$ junction is shown. The location of the MoS$_2$ layer is indicated by the dashed line. The simulation shows that a portion of the MoS$_2$ layer of length equal to a few hundred nanometers is photoexcited even underneath the metal.

A Finite Difference Time domain (FDTD) simulation of the optical excitation of the metal-MoS$_2$ junction by normally incident radiation is shown in Supplementary Figure 1. The location of the MoS$_2$ layer is indicated by the dashed line. The simulation shows that as a result of light diffraction from the edge of the metal contact, light scattering from the substrate, and plasmonic guidance, a portion of the MoS$_2$ layer of length equal to a few hundred nanometers is photoexcited even underneath the metal.

**Supplementary Note 2: Details on the Ultrafast Photoresponse of the Metal-MoS₂ Junction**

The nature of the ultrafast photoresponse of the metal-MoS$_2$ junction is discussed here. Figure 3(a) in the article depicts the band diagram of the metal-MoS$_2$ junction (plotted in the plane of the MoS$_2$ layer) after photoexcitation with an optical pulse. Given the Schottky barrier height of 100-300 meV [1], the width of the MoS$_2$ region near the metal with a non-zero lateral electric field is estimated to be to ∼100-300 nm [2]. Note that the lateral electric field right underneath the metal is expected to be very small. As a result of light diffraction from the edge of the metal contact, light scattering from the substrate, and plasmonic guidance, a portion of the MoS$_2$ layer of length equal to a few hundred nanometers is photoexcited even underneath the metal (see the discussion above). Assuming similar mobilities and diffusivities of electrons and holes in MoS$_2$, the photoexcited carriers, both free and bound (excitons), move, either by drift in the junction lateral electric field or by diffusion, less than ∼10 nm in 5 ps before they recombine and/or are captured by defects. The photoexcited carrier distributions therefore do not change significantly in space during their lifetime.

The ultrafast current response $I_2(t, \Delta t)$ of a short circuited junction in response to two time-delayed optical pulses is expected to be fairly complicated. In TPPC experiments the quantity measured is the time integral $\int I_2(t, \Delta t) \, dt \, (\propto V_c(\Delta t))$. The motion of the photoexcited electrons and holes in a short circuited junction causes capacitive (i.e. displacement) currents in the external circuit in order to keep the potential across the shorted junction from changing in accordance with the Ramo-Shockley theorem [3, 4]. However, if the photoexcited carriers recombine before they make it out into the circuit then the net contribution of the capacitive currents to the integral $\int I_2(t, \Delta t) \, dt$ is identically zero.

Photoexcited electrons and holes can be separated before they form excitons by the lateral electric field in the junction and this constitutes the standard drift current contribution to the detector short circuit current response $\int I_2(t, \Delta t) \, dt$. A photoexcited electron and a hole (free or belonging to an exciton) in the MoS$_2$ layer underneath the metal can also be separated at the metal-MoS$_2$ heterojunction. The hole can tunnel into the metal leaving behind the electron which is then swept by the lateral electric field to the opposite side of the junction. The electron can also tunnel into the metal leaving behind the hole which will then have a difficult time traversing the lateral field region (moving against the electric field) and making it to the opposite side of the junction. This argument shows that even if the probabilities of the electron and the hole tunneling into the metal are similar, the lateral field in the junction ensures that the process in which the

hole tunnels into the metal makes the dominant contribution to the short circuit current response $\int I_2(t, \Delta t)\, dt$. The experimentally measured sign of the photovoltage, and the photocurrent (see Figure 1(b) in the article), agrees with the above arguments.

The discussion above shows that the short circuit current response $\int I_2(t, \Delta t)\, dt$ is proportional to the junction lateral electric field strength, and to the time integral of the photoexcited free electron and hole densities as well as to the time integral of the bound (exciton) electron and hole densities. Assuming similar electron and hole mobilities, one may write,

$$\int I_2(t, \Delta t)\, dt \propto k_{\mathrm{f}} \int \left[ p'_{\mathrm{f}}(t, \Delta t) + n'_{\mathrm{f}}(t, \Delta t) \right]\, dt + k_{\mathrm{b}} \int \left[ p'_{\mathrm{b}}(t, \Delta t) + n'_{\mathrm{b}}(t, \Delta t) \right]\, dt \tag{1}$$

Here, $n'_{\mathrm{f/b}}(t, \Delta t)$ and $p'_{\mathrm{f/b}}(t, \Delta t)$ are the spatially-averaged free/bound (f/b) photoexcited electron and hole densities in the junction, respectively. Since photoexcited electrons and holes don't have time to move much before they recombine and/or are captured by defects, spatial dynamics of the carrier densities are not important. The constants $k_{\mathrm{f}}$ and $k_{\mathrm{b}}$ capture the difference in the relative contributions from free and bound carriers to the current response. If one assumes that $k_{\mathrm{f}} \approx k_{\mathrm{b}}$ then,

$$\int I_2(t, \Delta t)\, dt \propto \int \left[ p'(t, \Delta t) + n'(t, \Delta t) \right]\, dt \tag{2}$$

where $n'(t, \Delta t)$ and $p'(t, \Delta t)$ are the total photoexcited electron and hole densities in the junction, respectively, including carriers both free and bound (excitons). The assumption $k_{\mathrm{f}} \approx k_{\mathrm{b}}$ will hold if the short circuit current is dominated by the free and bound electrons and holes that get separated at the metal-MoS$_2$ heterojunction. Since the junction resistance $R_{\mathrm{j}}$ is expected to be largely determined by the transport across the metal-MoS$_2$ heterojunction rather than by the transport across the MoS$_2$ region, the assumption $k_{\mathrm{f}} \approx k_{\mathrm{b}}$ is a decent approximation if not an excellent one.

### Supplementary Note 3: High Frequency and Low Frequency Circuit Models of the Metal-MoS₂ Junction

A circuit model of the photodetector is shown in Supplementary Figure 2(a) and the resistances

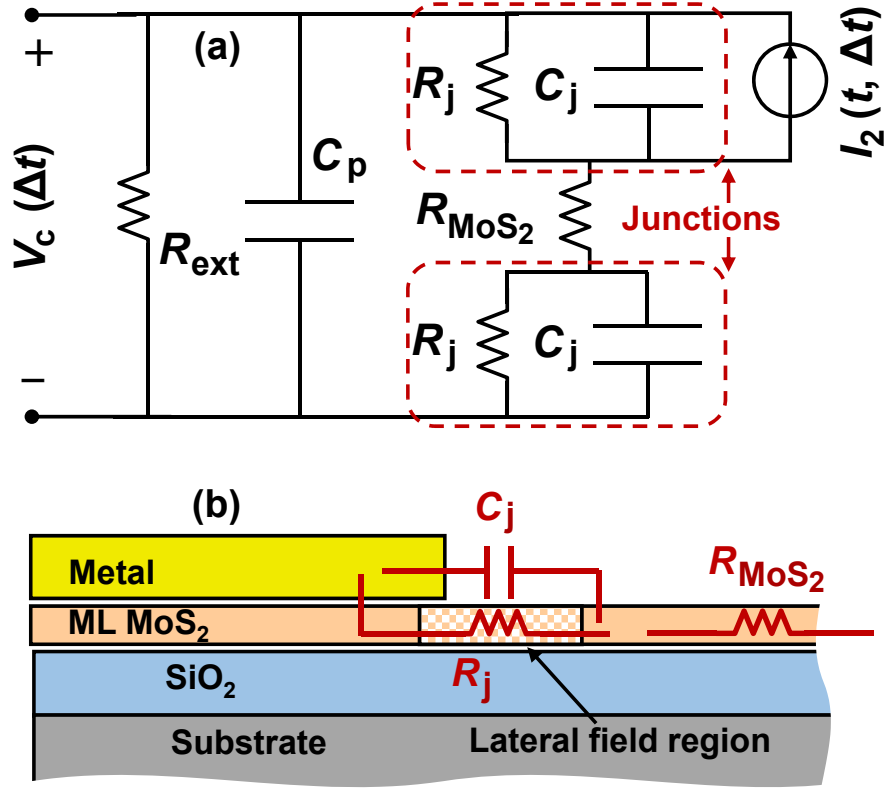

**Supplementary Figure 2: Circuit model of metal-MoS₂ photodetector.** A high-frequency circuit model of the photodetector is shown in (a). The resistances and capacitances associated with the metal-MoS₂ junction are depicted in (b). $R_j$ and $C_j$ are the resistance and the capacitance associated with the metal-MoS₂ junction. $R_{MoS_2}$ is the resistance of the undepleted MoS₂ region. $C_p$ is an external parasitic capacitance and $R_{ext}$ is the external circuit resistance (including the input resistance of the measurement instrument).

and capacitances associated with the metal-MoS₂ junction are shown in Supplementary Figure 2(b). The time-dependent short circuit current response of the photodetector to two optical pulses separated by time $\Delta t$ is $I_2(t, \Delta t)$. The short circuit current response $I_2(t, \Delta t)$ represents the photocurrent measured if the illuminated junction were shorted. In our experiments, the quantity measured is the DC value $V_c(\Delta t)$ of the open circuit voltage. Assuming periodic excitation of the

device with the optical pulses, $V_c(\Delta t)$ can be written as,

$$V_c(\Delta t) = \frac{1}{T_R}\frac{R_j R_{ext}}{R_d + R_{ext}}\int I_2(t, \Delta t)dt \approx \frac{R_j}{T_R}\int I_2(t, \Delta t)dt \quad (3)$$

Here, the total device resistance $R_d$ equals $2R_j + R_{MoS_2}$, $T_R$ is the period of the optical excitation, and the time integrals above are performed over one complete period. The approximate equality above follows from the fact that in our experiments, $R_{ex} >> R_d$. Note that all the capacitances drop out in the expression for $V_c(\Delta t)$. Therefore, one can use the low-frequency circuit model shown in Figure 1(c)(in the article) when calculating $V_c(\Delta t)$.

It is instructful to determine whether the intrinsic device resistances and capacitances could fundamentally limit the high speed performance. The frequency dependent small-signal open circuit voltage response and short circuit current response of the detector can also be evaluated using the circuit shown in Figure 2(a) under the assumption that $I_2(t) = \text{Re}\left\{i_2(\omega)e^{-i\omega t}\right\}$. The open circuit voltage response is (assuming $R_{ext} = \infty$),

$$\frac{v_c(\omega)}{i_2(\omega)} = \frac{R_j}{1 - i\omega C_j R_j - i\omega C_p(2R_j + R_{MoS_2}) + (i\omega)^2 C_p C_j R_j R_{MoS_2}} \quad (4)$$

If one ignores the parasitic capacitance $C_p$ then the circuit bandwidth is set by the time constant $R_j C_j$. The junction resistance $R_j$ is dominated by the metal-semcionductor contact resistance. In the case of $MoS_2$, the contact resistance values are in the 1-10 k$\Omega$-$\mu$m range at room temperature [8]. Because of the 2D nature of the metal-semiconductor junction, the junction capacitance $C_j$ is very small and entirely due to the fringing fields. For a $\sim$50 nm thick metal contact layer, $C_j$ is estimated to be less than .03 fF/$\mu$m [2]. Therefore, the relevant time constant is estimated to be shorter than a picosecond. The short circuit current response is (assuming $R_{ext} = 0$),

$$\frac{i_{ext}(\omega)}{i_2(\omega)} = \frac{R_j}{2R_j + R_{MoS_2} - i\omega C_j R_j R_{MoS_2}} \quad (5)$$

In this case, the circuit bandwidth is set by the time constant $C_j R_j R_{MoS_2}/(2R_j + R_{MoS_2})$. If $R_j << R_{MoS_2}$, as would be the case if the doping in the $MoS_2$ sheet is small, then the time constant equals $\sim R_j C_j$, which is the same as for the open circuit voltage response. If on the other hand $R_j >> R_{MoS_2}$, then the time constant equals $\sim 0.5R_{MoS_2}C_j$. Assuming an electron doping of $\sim 2 \times 10^{12}$ cm$^{-2}$ (as in our devices) and a modest electron mobility of $\sim$20 cm$^2$/V-s (as in our devices), and a device length of 1 $\mu$m, the value of $R_{MoS_2}$ comes out to be $\sim$150 k$\Omega$-$\mu$m and the time constant comes out to be $\sim$2.25 ps. Finally, the capacitance $C_p$ could come from the fringing fields between the two metal contacts and therefore its effect on the open circuit voltage response ought to be considered. For example, consider $\sim$50 nm thick metal contact layers that are one

micron apart. The capacitance $C_\mathrm{p}$ is estimated to be less than .02 fF/$\mu$m [9]. The relevant time constant is $C_\mathrm{p}(2R_\mathrm{j} + R_{\mathrm{MoS_2}})$ and, assuming $R_\mathrm{j} << R_{\mathrm{MoS_2}}$, the time constant is found to be $\sim$3.0 ps. Therefore, in all the cases the intrinsic device resistances and capacitances are not expected to fundamentally limit the speed of operation of the detectors considered in this work.

**Supplementary Note 4: Gate Bias Dependence of the Photoresponse**

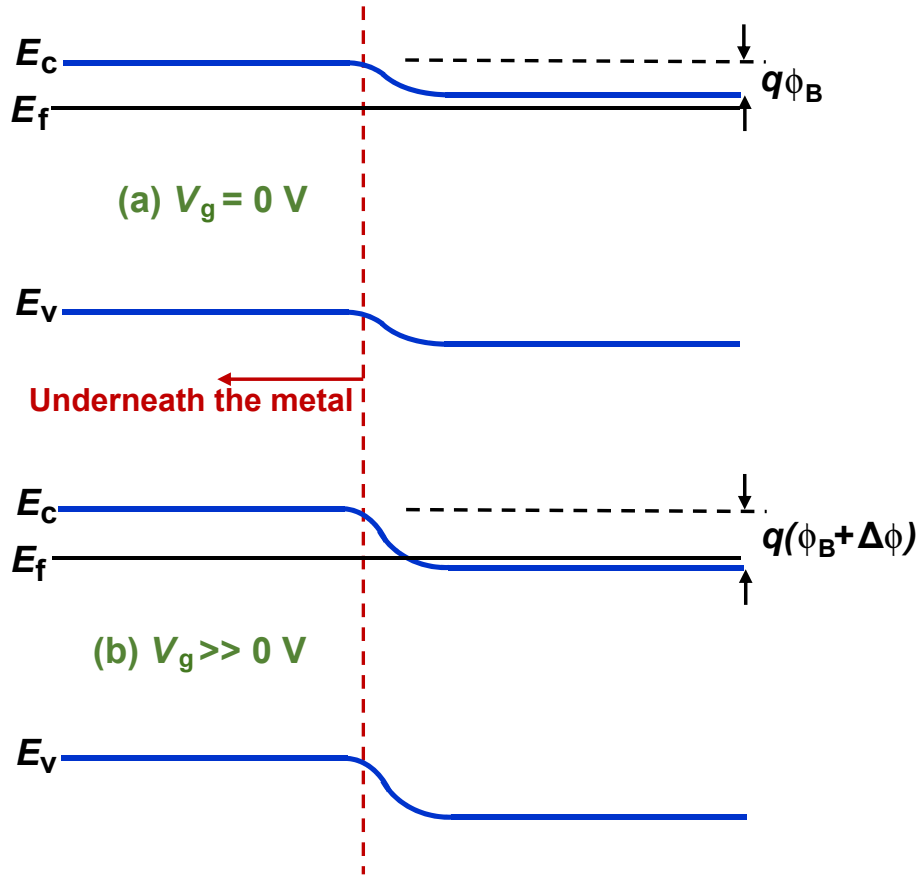

**Supplementary Figure** 3: **Band diagram of metal-MoS₂ junction.** The effect of the gate voltage $V_g$ on the energy band diagram of the the metal-MoS₂ junction is depicted when (a) $V_g = 0$ V, and (b) $V_g >> 0$ V. The in-plane electric field increases near the metal junction with an increase in the gate voltage.

The effect of the gate voltage on the photoresponse and the TPPC measurements is discussed here. Increasing the back gate voltage increases the electron density in the MoS₂ layer and raises the Fermi level with respect to the conduction band edge. However, in the region of the MoS₂ layer which is underneath the metal the electron density as well as the position of the Fermi level with respect to the conduction band edge remains unchanged. This is because the metal on top screens all the extra charges on the gate as the gate voltage is increased. Consequently, the in-plane lateral electric field in the MoS₂ layer increases near the metal junction with an increase in the gate voltage, as depicted in Supplementary Figure 3. The exact magnitude of the increase in the

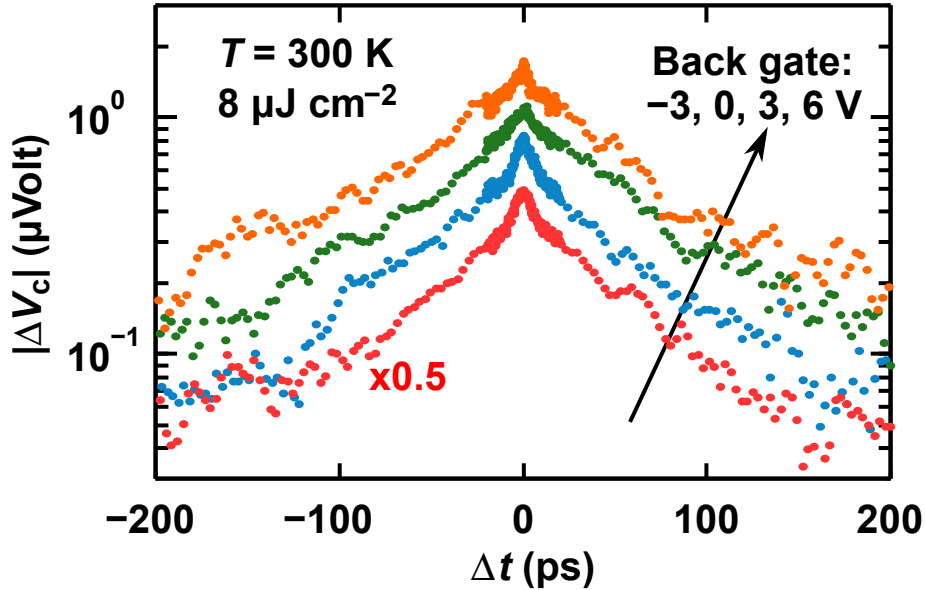

**Supplementary Figure 4: Gated TPPC experiment results.** The measured two-pulse photovoltage correlation (TPPC) signal $|\Delta V_c(\Delta t)|$ is plotted as a function of the time delay $\Delta t$ between the pulses for different gate bias values: -3, 0, 3, 6 V. $T = 300$K. The pump fluence is 8 $\mu$J cm$^{-2}$. As in Figure 2(b,c) in the article, two distinct time scales are observed in the dynamics and these time scales are largely independent of the gate bias.

in-plane electric field with the gate voltage is hard to predict since the result could depend on the degree of Fermi level pinning on defect states within the bandgap in MoS$_2$. Nevertheless, one would expect the measured photoresponse to also increase with the gate voltage since, as argued in this paper, the photoresponse is proportional to the in-plane electric field [5].

Supplementary Figure 4 shows the measured $|\Delta V_c(\Delta t)|$ plotted as a function of the time delay $\Delta t$ between the pulses for different gate bias values: -3, 0, 3, 6 V. $T = 300$K. The pump fluence is 8 $\mu$J cm$^{-2}$. As in Figure 2(b,c) in the article, two distinct time scales are observed in the dynamics and, within the accuracy of our measurements, these time scales are largely independent of the gate bias. As expected from our model, the overall signal level increases with the gate bias.

**Supplementary Note 5: Theoretical Model for Carrier Capture and Recombination via Auger Scattering**

We use the model for carrier capture by defects via Auger scattering in $MoS_2$ [6, 7] to model our experimental TPPC results. The model assumes carrier capture by two different defect levels, one fast ($f$) and one slow ($s$). Keeping only the dominant Auger capture processes in an n-doped sample, and ignoring carrier emission processes for simplicity, the rate equations for the carrier densities and defect occupation probabilities with photoexcitation by two time-delayed optical pulses can be written as follows [6],

$$\frac{dn(t, \Delta t)}{dt} = -A_f n_f (1 - F_f) n^2 - A_s n_s (1 - F_s) n^2 + g I_p(t) + g I_p(t - \Delta t) \tag{6}$$

$$\frac{dp(t, \Delta t)}{dt} = -B_f n_f F_f np - B_s n_s F_s np + g I(t) + g I(t - \Delta t) \tag{7}$$

$$n_{f/s} \frac{dF_{f/s}(t, \Delta t)}{dt} = A_{f/s} n_{f/s} n^2 (1 - F_{f/s}) - B_{f/s} n_{f/s} np F_{f/s} \tag{8}$$

Here, $n(t, \Delta t)$ and $p(t, \Delta t)$ are the total electron and hole densities, respectively, including both free and bound (excitons) carriers and $n(t, \Delta t) = n_o + n'(t, \Delta t)$, where $n_o$ is the doping density. $F_{f/s}$ are the defect occupation probabilities. $A_{f/s}$ and $B_{f/s}$ are the Auger capture rate constants for electrons and holes, respectively. $n_{f/s}$ are the defect densities. $I_p(t)$ is the optical pulse intensity ($\mu W\ cm^{-2}$) and $g$, determined from the measured $MoS_2$ optical absorption at the wavelength of the optical pulse excitation, equals $\sim 2.5 \times 10^{11}$ $(\mu J)^{-1}$ and corresponds to around 11% absorption in monolayer $MoS_2$ on oxide at 452 nm wavelength [6]. In our n-doped sample, the defects are assumed to be fully occupied before photoexcitation. The above rate equations can be solved in time and the resulting photoexcited carrier densities integrated in time to yield the measured photovoltage correlation signal (up to a multiplicative constant). The results are shown in Supplementary Figure 5(a-c) and compared with the measurement results.

The values of the fitting parameters used in the theoretical model to fit the experimental data (see Figure 3 in the article) are listed in Supplementary Table 1. These values are almost identical to the values extracted from direct optical pump-probe measurements of the carrier dynamics in monolayer $MoS_2$ [6]. The value of the doping density, $n_o \sim 8 \times 10^{11}\ cm^{-2}$, needed to obtain a good match with the experiments is much smaller than the doping density determined from electrical transport measurements. This difference is attributed to the fact that the carrier density in $MoS_2$ near the metal contact is indeed much smaller than in the bulk of the device (see the energy band diagram in Figure 3(a) in the article).

In the simulations, we assumed, for simplicity, that the defect occupation probability before

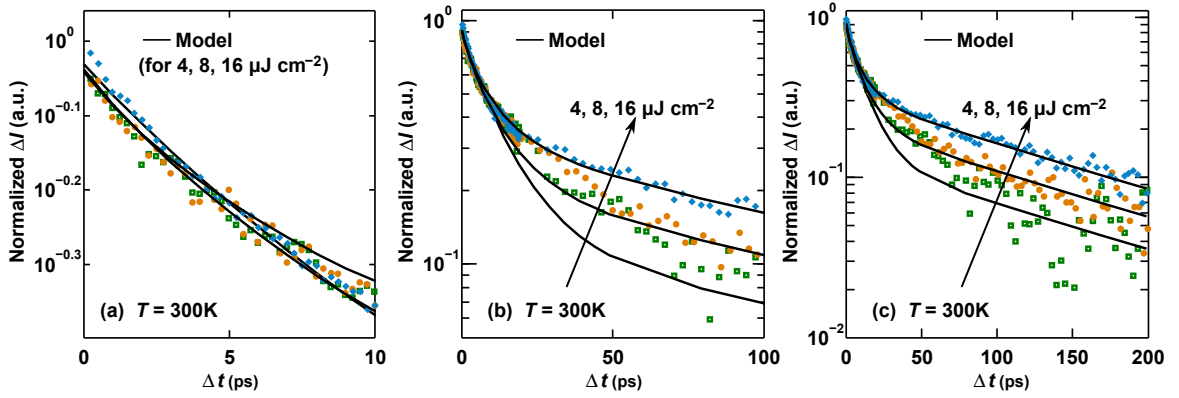

**Supplementary Figure** 5: **Details of the theoretical fitting.** The measured (symbols) and computed (solid lines) photovoltage correlation signals $|\Delta V_c(\Delta t)|$, normalized to the maximum value, are plotted as a function of the time delay $\Delta t$ for different pump fluence values: 4, 8, and 16 $\mu$J cm$^{-2}$ and for different temporal resolutions (a-c). $T = 300$K. Note that at short time scales the measured transients are not exactly decaying exponentials. The carrier capture model reproduces all the time scales observed in the measurements over the entire range of the pump fluence values used.

| | |
|---|---|
| $B_f n_f$ | $0.73 \pm 0.05$ cm$^2$ s$^{-1}$ |
| $B_s n_s$ | $2.79 \pm 1 \times 10^{-2}$ cm$^2$ s$^{-1}$ |
| $A_s$ | $9.5 \pm 1 \times 10^{-15}$ cm$^4$ s$^{-1}$ |
| $A_f$ | $(1.0 \pm 0.2)B_f$ |
| $n_{f/s}$ | $5.0 \times 10^{12}$ cm$^{-2}$ |
| $n_o$ | $8 \times 10^{11}$ cm$^{-2}$ |

**Supplementary Table** I: **Fitting parameters.** Parameter values used in the simulations to fit the photovoltage correlation data.

photoexcitation is unity. This assumption might not always hold. The defect occupation probability could be a function of the temperature or the gate bias and this could have observable consequences. If the occupation probability of the slow defects before photoexcitation is smaller at higher temperatures, which is plausible, then the ratio of $|\Delta V_c(\Delta t = 0)|$ to $|\Delta V_c(\Delta t)|$, when $\Delta t$ is at the boundary between the slow and fast time constant regions, will be larger at higher temperatures since fewer photoexcited holes would have been captured by the slow defects during the initial fast transient and, consequently, fewer photoexcited electrons would have remained in the conduction band after the fast transient is over. This could explain the small decrease in the ratio of $|\Delta V_c(\Delta t = 0)|$ to $|\Delta V_c(\Delta t)|$ at the boundary between the slow and fast time constant

regions observed in our measurements at the lower temperature in Figure 2(b) in the article. Note also that the ratio of $|\Delta V_c(\Delta t = 0)|$ to $|\Delta V_c(\Delta t)|$, when $\Delta t$ is at the boundary between the slow and fast time constant regions, is also smaller at more positive gate bias values (see Supplementary Figure 4). This trend is similar to the trend observed in $|\Delta V_c(\Delta t)|$ when going to lower substrate temperatures, and, as before, we attribute this to a larger initial occupancy of the slow defect states just before photoexcitation at more positive gate bias values.

--------

**SUPPLEMENTARY REFERENCES**



\end{document}